\documentclass{article}
\usepackage{ijcai24}

\usepackage{times}
\usepackage{soul}
\usepackage{url}
\usepackage[hidelinks]{hyperref}
\usepackage[utf8]{inputenc}
\usepackage[small]{caption}
\usepackage{graphicx}
\usepackage{amsmath}
\usepackage{amsthm}
\usepackage{booktabs}
\usepackage{algorithm}
\usepackage{algorithmic}
\usepackage[switch]{lineno}
\usepackage{subfigure}

\usepackage{multirow}
\usepackage{multicol}
\usepackage{color, colortbl}
\usepackage[cmyk]{xcolor}
\definecolor{LightCyan}{rgb}{0.41,0.91,0.92}
\colorlet{jason}{black}
\colorlet{jie}{black}
\colorlet{binxiao}{black}
\colorlet{nwong}{black}

\makeatletter
\renewcommand\@fnsymbol[1]{}
\makeatother

\title{Hundred-Kilobyte Lookup Tables for Efficient Single-Image Super-Resolution}

\author{
Binxiao Huang\thanks{$\dagger$: Equal contribution}$^{1 \dagger}$
\and
Jason Chun Lok Li$^{1 \dagger}$
\and
Jie Ran$^1$ 
\and \\
Boyu Li$^1$
\and
Jiajun Zhou$^1$
\and
Dahai Yu$^2$
\and
Ngai Wong$^1$
\\
\affiliations
$^1$ The University of Hong Kong \\
$^2$ TCL Corporate Research\\
\emails
\{huangbx7, jasonlcl\}@connect.hku.hk, \{jieran, liboyu, jjzhou, nwong\}@eee.hku.hk,  dahai.yu@tcl.com\\
}

\begin{document}

\maketitle

\begin{abstract}
Conventional super-resolution (SR) schemes make heavy use of convolutional neural networks (CNNs), which involve intensive multiply-accumulate (MAC) operations, and require specialized hardware such as graphics processing units. This contradicts the regime of edge AI that often runs on devices strained by power, computing, and storage resources. Such a challenge has motivated a series of lookup table (LUT)-based SR schemes that employ simple LUT readout and largely elude CNN computation. Nonetheless, the multi-megabyte LUTs in existing methods still prohibit on-chip storage and necessitate off-chip memory transport. This work tackles this storage hurdle and innovates hundred-kilobyte LUT (HKLUT) models amenable to on-chip cache. Utilizing an asymmetric two-branch multistage network coupled with a suite of specialized kernel patterns, HKLUT demonstrates an uncompromising performance and superior hardware efficiency over existing LUT schemes. Our implementation is publicly available at: \url{https://github.com/jasonli0707/hklut}.
\end{abstract}

\section{Introduction}
Single image super-resolution (SISR) is a long-standing problem in computer vision that involves generating a high-resolution (HR) image from a low-resolution (LR) image. The goal is to recover high-frequency details that are lost in a downsampled image. Classical non-deep-learning methods can be divided into two main categories: interpolation and sparse coding. Interpolation approaches~\cite{keys1981cubic,kirkland2010bilinear}, such as bicubic interpolation, are simple and computationally efficient. They involve upsampling the LR image using a fixed kernel, e.g., a bicubic filter. However, they often suffer from limited generalizability and produce blurry images. Sparse-coding approaches~\cite{timofte2013anchored,timofte2015a+}, on the other hand, can achieve better results than the former but are computationally expensive and require hand-crafted features.

\begin{figure}[th]
    \centering
    \includegraphics[width=1\columnwidth]{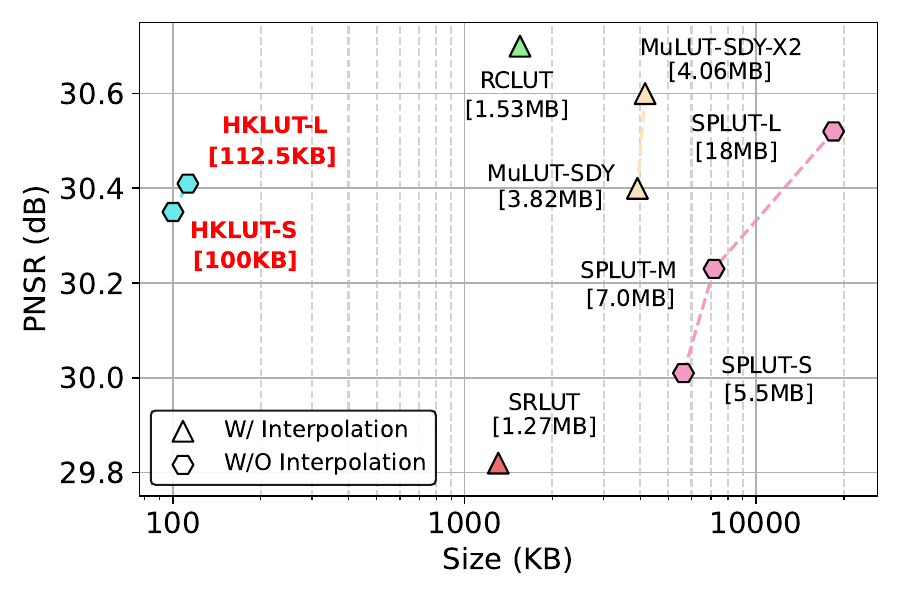}
    \caption{PSNR (dataset: Set5) vs. model size: HKLUTs are the \textit{first} hundred-KB series of LUT-based SISR models targeting the sub-1MB regime (NB. $x$-axis is in log scale).}\vspace{-12pt}
    \label{fig:psnr_vs_latency}
\end{figure}

Like many other problems in computer vision, deep learning has revolutionized SISR. The mapping between LR and HR images can be learned directly from data, often resulting in super-resolved images of much better quality than classical methods. However, deep neural networks (DNNs) require a large number of floating-point operations, resulting in a high power budget. Recently, researchers have explored new schemes that combine lookup tables (LUT) with deep-learning SISR~\cite{jo2021srlut,li2022mulut,ma2022splut}. Specifically, this approach pre-computes all possible input-output pairs of a deep SR network and stores them into an LUT for retrieval, thus skipping expensive computation during inference. As a result, LUT-based methods achieve inference speeds comparable to interpolation methods~\cite{jo2021srlut}, but produce results with a learning-based quality. Moreover, LUT-based approaches are agnostic to software and hardware platforms, making them much more versatile than DNNs which require dedicated software and hardware support. Nonetheless, recent LUT-based SR schemes~\cite{li2022mulut,ma2022splut} mainly focus on improving performance by enlarging the receptive field (RF), while overlooking the memory constraints. This contradicts the trending edge AI that often runs on edge devices strained by limited resources, e.g., the popular Ultra96 FPGA board has $<1$MB on-chip memory. In fact, existing LUT-based schemes generally require multi-megabyte LUTs where off-chip storage is inevitable. In addition, existing LUT-based approaches may still require expensive interpolation and/or floating-point operations, resulting in higher peak memory and energy costs. This highlights the challenge as well as the need for significant LUT size reduction and better architecture design to address the storage and power issues, while upkeeping performance.

To this end, this work systematically explores the relationship between the number of input pixels, RF, and SR performance. Our study leads to the design of a family of models called Hundred-Kilobyte Look-up Tables (HKLUTs) which offer a compelling balance between memory and performance. To eliminate the time-consuming and energy-intensive interpolation, a two-branch architecture as in SPLUT~\cite{ma2022splut} is employed. Yet instead of its symmetric branches~\cite{ma2022splut}, we exploit the idea of effective receptive field (ERF)~\cite{luo2016understanding} in different branches and reveal a larger RF is only required for the most-significant-bit (MSB) branch representing contextual semantic information, whereas a much smaller RF is sufficient for processing the least-significant-bit (LSB) branch.

This key insight leads to an innovative asymmetric parallel structure, which largely reduces the storage space by nearly half without sacrificing performance. Next, we extend our model to multistage by cascading LUTs to enlarge the RF. Our proposed architecture, in contrast to prior works, allows information exchange between two branches during stage transition, resulting in an uncompromising performance. Moreover, we are the first to utilize progressive upsampling on LUT-based approaches, which further reduces the LUT size by half while improving the final performance. Compared to other state-of-the-art (SOTA) LUT-based schemes, HKLUTs deliver competitive performance while being $>10\times$ smaller in size (cf. Fig.~\ref{fig:psnr_vs_latency}). The main contributions of this paper are threefold:
\begin{itemize}
    \item We systematically study the relationship between RF, the number of input pixels, and performance in the SR task, leading to our design of extra lightweight LUTs.
    \item We investigate the ERFs of two branches and propose an asymmetric parallel structure to reduce storage by $\approx 2\times$ without sacrificing performance.
    \item A novel multistage architecture is proposed which adopts a progressive upsampling strategy and enables the communication between the two branches during stage transition. This leads to an improved performance together with another $2\times$ reduction in size.
\end{itemize}
By seamlessly integrating the above techniques, HKLUTs constitute the smallest LUT-based SISR schemes in the literature offering a highly competitive performance.

\section{Related Work}
The quest for high-quality SR on edge devices has fueled the interest and research of effective and efficient SISR schemes in recent years, broadly categorized into classical, deep learning and LUT-based Approaches.

\noindent {\bf Classical Approaches.} Basic linear interpolation methods~\cite{keys1981cubic,kirkland2010bilinear}, though simple and efficient, are limited in their abilities to adapt to the image content and often result in blurry images. Early example-based work~\cite{timofte2013anchored,timofte2015a+} learned sparse dictionaries over pixels or patches to build a compact representation between LR and HR pairs. During inference, the HR output is obtained using the pre-computed projection matrix with the input features. However, it remains time-consuming for sparse coding models to compute the sparse representations. 

\noindent {\bf Deep Learning Approaches.} Deep learning methods have shown promising results in reconstructing image details for accurate super-resolution. Many deep learning SR works have been developed to reduce computational burdens by carefully adjusting the network structures, including, 
ESPCN~\cite{shi2016real}, CARN~\cite{ahn2018fast}, RRDB~\cite{wang2018esrgan} and IMDN~\cite{hui2019lightweight}. LapSRN~\cite{lai2017deep} and its extension MS-LapSRN~\cite{lai2018fast} employ Laplacian Pyramid to generate multi-scale predictions and progressively reconstruct HR images in multiple steps. FMEN~\cite{du2022fast} shined in the NTIRE 2022 challenge with its low memory cost and short runtime through the enhanced residual block and high-frequency attention module. Although some models can be executed in real-time on GPUs, it is impractical to deploy them on edge devices with constrained hardware resources.

\noindent {\bf LUT-based Approaches.} While deep learning-based methods can produce impressive results, they often require heavy computations and large storage. To accelerate the inference of DNNs, dedicated software and hardware platforms such as CUDA and GPUs are required. This poses challenges for cost-sensitive and resource-restrictive edge devices such as smartphones and TVs. To address this, SRLUT~\cite{jo2021srlut} fuses DNN and LUT for SR applications by pre-computing and caching input-output pairs of a pre-trained SR network into a LUT for retrieval at test time. This approach eliminates expensive computation during DNN inference and achieves similar speed but better SR results than classical approaches.  Two independent works around the same time, namely, MuLUT and SPLUT~\cite{li2022mulut,ma2022splut}, propose to expand the RF covered by cascading LUTs. MuLUT proposes the use of multiple complementary LUTs together with the rotation ensemble to solve the intrinsic limitation of a small RF in SRLUT. 
However, MuLUT preserves the costly interpolation step of SRLUT. On the other hand, SPLUT~\cite{ma2022splut} exploits a novel parallel framework where the quantization loss is compensated by using two parallel branches that separately handle the most significant 4 bits and least significant 4 bits. This eliminates the time-consuming interpolation step for a faster inference speed. SPLUT also suggests enlarging the RF through feature aggregation along the horizontal and vertical directions. In addition, the Reconstructed Convolution (RC) module proposed in ~\cite{liu2023rclut} can be transformed into 1D LUTs to efficiently expand the RF, thus improving performance. However, while these methods enhance performance compared to SRLUT, they require more storage space ($>1$MB), making them less suitable for edge devices.

\section{Development of HKLUT}
\subsection{Reduced Number of Input Pixels}
The storage size of a LUT can be calculated using the formula $v^n \times r^2$, where $v$, $n$, and $r$ are the number of quantization levels, number of input pixels, and upscaling factor, respectively. The first term $v^n$ determines the number of entries, whereas the second term $r^2$ refers to the length of the output vector of each LUT. It is obvious that reducing the number of input pixels $n$ can result in an exponential reduction in storage space. Previous studies have primarily focused on expanding the RF by utilizing a fixed number of pixels (say, 4) to improve the final performance. SPLUT~\cite{ma2022splut} uses feature aggregation to expand the RF, namely, the output feature maps generated from the previous LUT are divided into groups along the channel dimension and then padded horizontally or vertically. The padded feature maps are then added to generate a single-channel feature map, which serves as the input for subsequent LUTs. However, in order to generate multi-channel feature maps, the size of intermediate LUTs grows linearly with the number of output channels. This goes against our primary goal, which is to reduce storage space. Subsequently, we deliberately omit feature aggregation. On the other hand, SRLUT~\cite{jo2021srlut} and MuLUT~\cite{li2022mulut} employ rotation ensemble to increase the area being covered. The final prediction $\hat{y}_i$ is computed as follows:
\begin{equation}
\label{eqt:rotation_ensemble}
\hat{y}_i = \frac{1}{N} \frac{1}{M_k}\sum_{k=0}^{N}\sum_{j=0}^{M_k}R^{-1}_j(LUT_k(R_j(x_i))
\end{equation}
where $x_i$ is the LR input, $LUT_k$ is the $k$th LUT, $R_j$ is the $j$th rotation operation, $N$ is the number of kernels (or LUTs), and $M_k$ is the number of rotation operations for the corresponding kernel. We remark that rotation ensemble, unlike feature aggregation, does not require extra storage.

\begin{figure}[th]
    \centering
    \includegraphics[width=0.92\columnwidth]{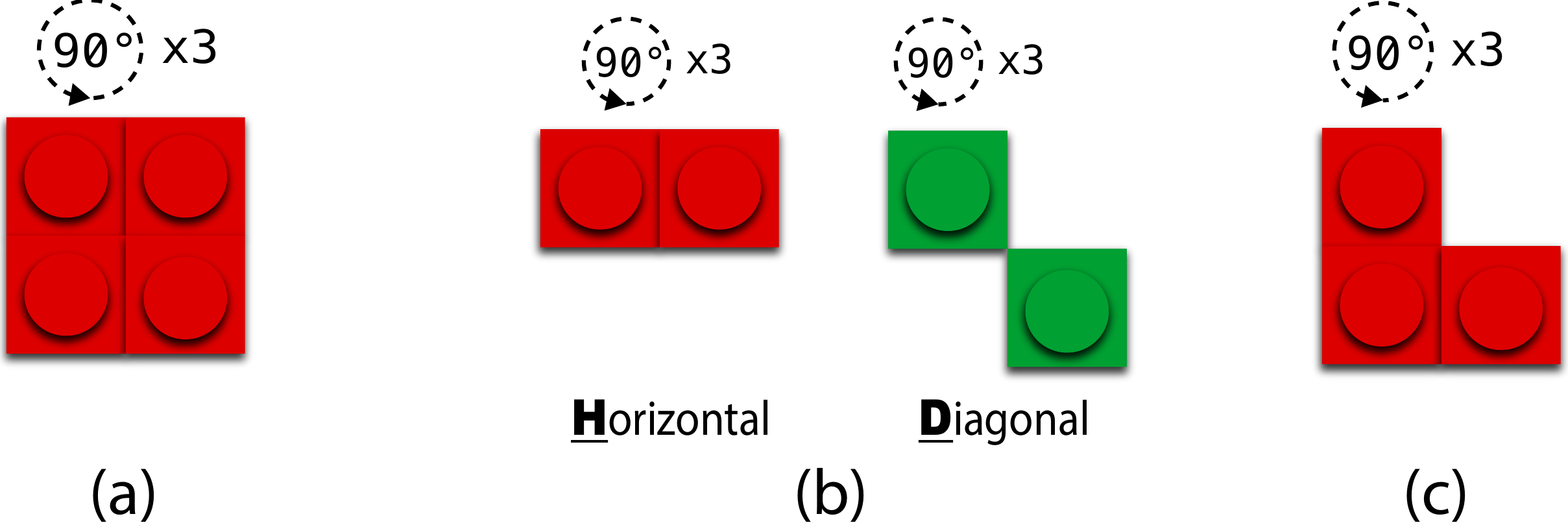}
    \caption{Three types of kernels covering a $3\times3$ RF with rotation ensemble via different numbers of input pixels. (a) SRLUT (b) HDLUT (c) LLUT.}
    \label{fig:effect_of_pixels}
\end{figure}

\begin{table}[ht]
\begin{center}
\resizebox{0.9\columnwidth}{!}{
\begin{tabular}{|l|cccc|}
\hline
Method & RF & \# Pixels & LUT size & PSNR(dB) \\
\hline\hline
SRNet & $3 \times 3$     & 4 & 1.27 MB & \textbf{29.87} \\
LNet & $3 \times 3$     & 3 & 76.77 KB & 29.67 \\
HDNet & $3 \times 3$   & 2 & \textbf{9.03 KB} & 29.45\\
\hline
\end{tabular}%
}
\end{center}
\caption{Effect of the number of input pixels on PSNR, evaluated on Set5 dataset. ``Net'' is used to distinguish full-precision CNNs from quantized LUTs.}
\label{tab:effect_of_pixels}
\end{table}

\begin{figure}[th]
    \centering
    \includegraphics[width=0.85\columnwidth]{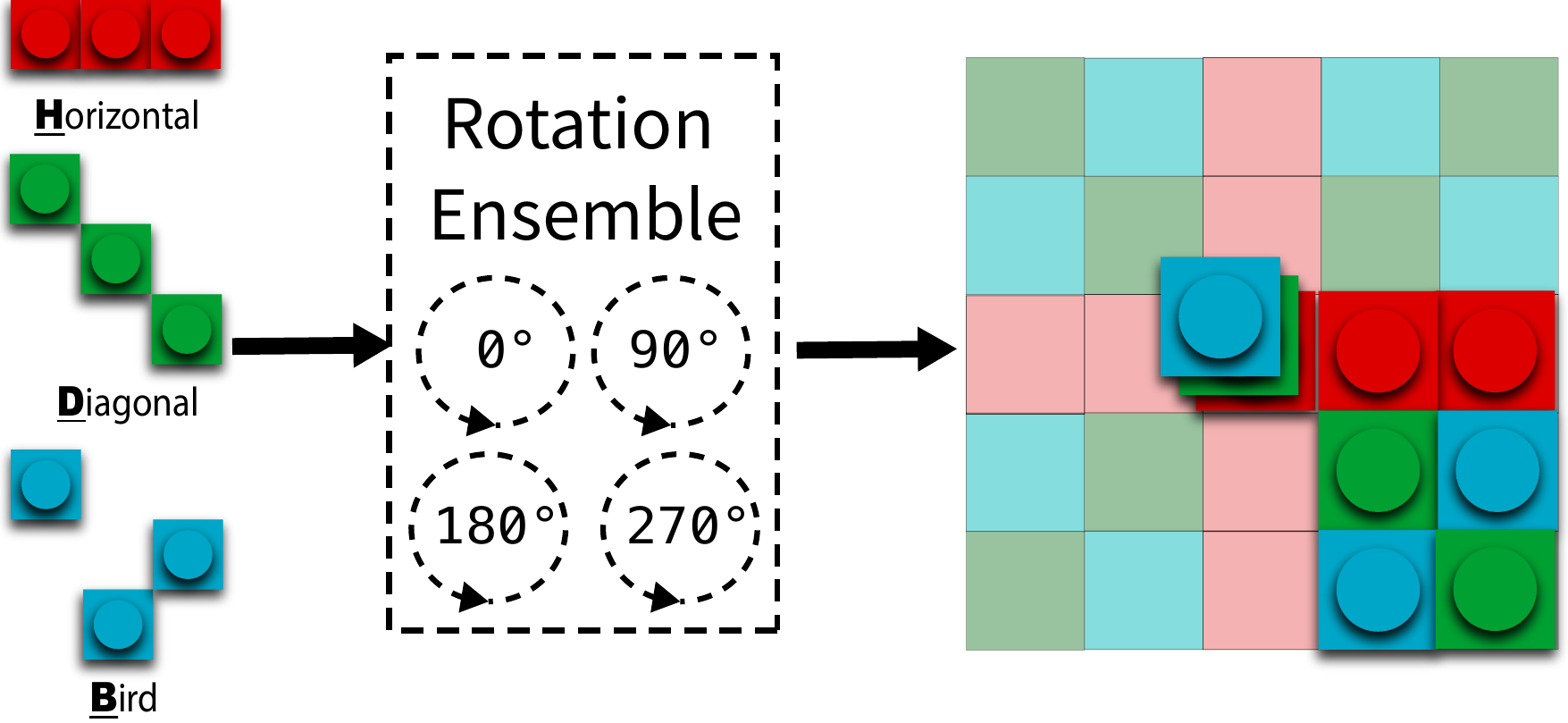}
    \caption{With rotation ensemble, HDBLUT can cover a $5\times5$ area with 3 three-pixel kernels. On the right, each square represents a single pixel, with a color indicating which kernel covers it.}\vspace{-12pt}
    \label{fig:hklut_rotation_ensemble}
\end{figure}

An important aspect that has been overlooked by previous works is that achieving the same RF is indeed feasible \textit{with fewer pixels}. To fill in this gap, we design LLUT and HDLUT (named after their input-pixel shapes, cf. Fig.~\ref{fig:effect_of_pixels}) to explore the relationship between the number of input pixels and the final PSNR subject to a prescribed RF size. Specifically, LLUT and HDLUT are designed to completely cover a $3\times3$ area without any overlapping pixels, except for the pivot pixel, using a rotation ensemble. Table~\ref{tab:effect_of_pixels} shows the PSNR improves with an increasing number of input pixels, which is consistent with the intuition that having more pixels gathers more information, at the cost of a higher storage, viz. a 4-pixel LUT requires megabytes of storage, whereas the performance of the 2-pixel LUT degrades obviously. To balance storage and performance, we devise HDBLUT, which consists of three 3-pixel kernels. Using rotation ensemble, it covers an entire $5\times5$ RF without any overlap, except for the pivot pixel (Fig.~\ref{fig:hklut_rotation_ensemble}). In contrast to SRLUT and MuLUT, which have a $3\times3$ RF with 4 overlapping pixels and a $5\times5$ RF with 11 overlapping pixels, respectively, our designs, HDLUT and HDBLUT, aim to match the same RF with minimal input pixels. This achieves a near-optimal tradeoff between memory footprint and performance (a comparison of the RF of our designs vs others is in the Appendix). Using HDBLUT and HDLUT, along with the asymmetric parallel structure discussed in the sequel, we are able to outperform SRLUT with $>6\times$ less storage.

\subsection{Asymmetric Parallel Structure}
\label{subsec:asymmetric_parallel_structure}
Previous studies have attempted to reduce the size of LUTs by limiting the number of quantization levels to 17 and using interpolation to recover the quantization loss~\cite{jo2021srlut,li2022mulut}. However, the interpolation stage is time-consuming and energy-intensive, hindering the practicality on edge devices. To address this, SPLUT~\cite{ma2022splut} proposes a parallel branch structure that avoids the need for interpolation. The structure divides the 8-bit input image into its most and least significant bits, viz. MSB and LSB branches each with 4 bits (viz. 16 levels) of information. The final SR image is obtained by fusing the results of the two branches at the last stage. While this approach gives promising results, it overlooks an important fact, namely, the effective receptive fields (ERFs) of MSB and LSB branches are completely different as they represent different kinds of information. Therefore, using the same sets of LUTs for both MSB and LSB branches might not be an optimal choice. 

\begin{figure}[t]
    \centering
    \includegraphics[width=1\columnwidth]{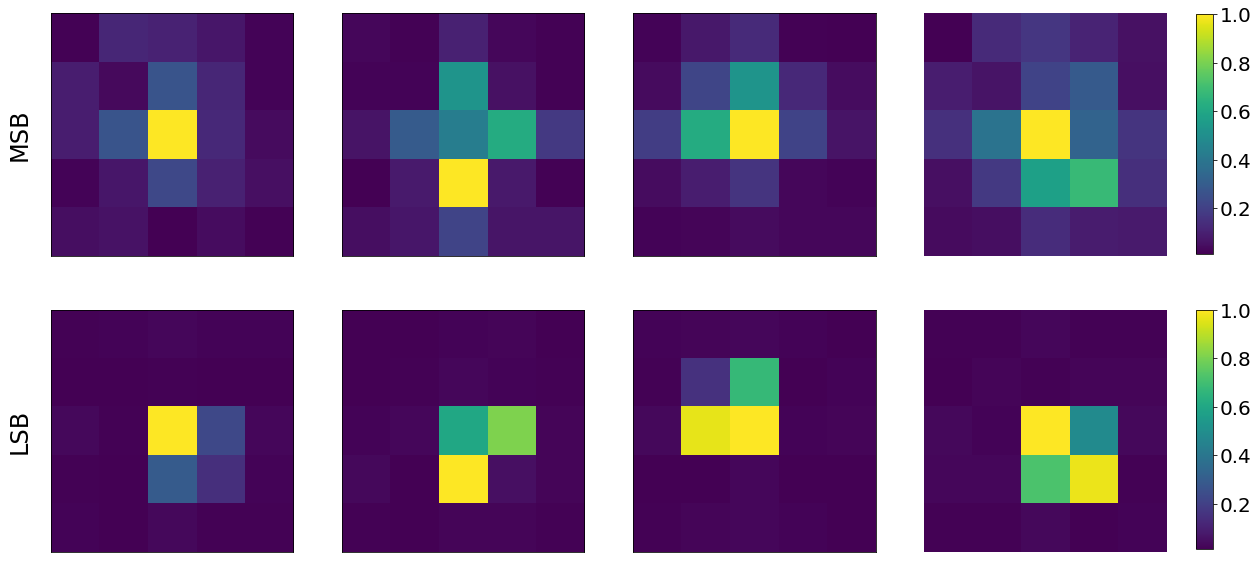}
    \caption{Four randomly selected effective receptive fields (ERFs). \textbf{Top:} Most significant 4 bits. \textbf{Bottom:} Least significant 4 bits. The brightness denotes model's sensitivity to that pixel, justifying the assymetric $5\times5$ and $3\times3$ kernels for respective branches.}\vspace{-12pt}
    \label{fig:msb_vs_lsb}
\end{figure}

We trained two identical SRNets, using a $5\times5$ kernel for both MSB and LSB branches. The results from the two branches are summed and gradients are computed using mean-squared error (MSE) loss. Fig.~\ref{fig:msb_vs_lsb} shows the ERF, in terms of normalized gradients obtained from randomly selected $4\times4$ regions of the SR result with respect to inputs, using the baby image from the Set5 dataset (cf. Fig.~\ref{fig:overall_architecture}) (more demonstrations can be found in the Appendix). Although both MSB and LSB are processed under the same theoretical RF (i.e., $5\times5$), the results show that the MSB branch, which captures the high-level contextual semantic information, has a larger ERF. On the other hand, a smaller ERF is sufficient for the LSB branch to describe fine details. This observation justifies an asymmetric parallel structure that exploits the differences between MSB and LSB to save storage space. By using a set of 3-pixel LUTs (i.e., HDBLUT) corresponding to CNNs with a $5\times5$ RF for the MSB branch, and a set of 2-pixel LUTs (i.e., HDLUT) corresponding to CNNs with a $3\times3$ RF for the LSB branch, we can nearly halve the storage space without compromising performance (cf. Table~\ref{tab:effect_of_asymmetric_parallel}).

\begin{figure*}[th]
    \centering
    \includegraphics[width=1.9\columnwidth]{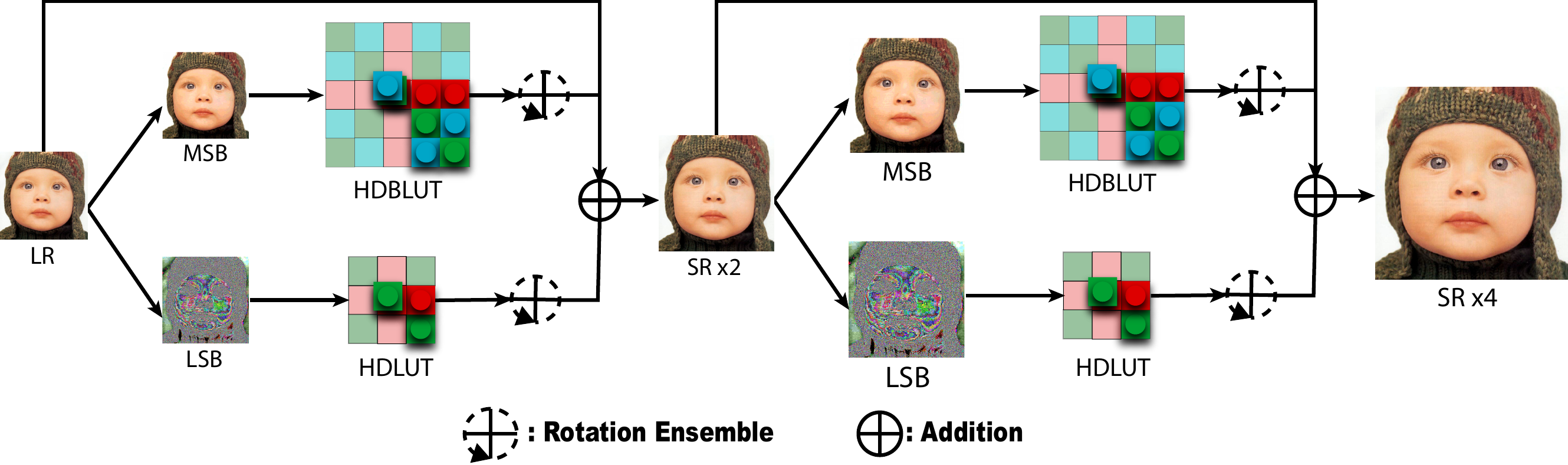}
    \caption{Overall architecture of the proposed method (HKLUT-S).}
    \label{fig:overall_architecture}
\end{figure*}

\subsection{Multistage}
\label{subsec:multistage}
\textcolor{binxiao}{LUT-based methods~\cite{ma2022splut,li2022mulut} have demonstrated the efficacy of cascaded stages to enlarge the RF for better performance. SPLUT~\cite{ma2022splut} splits the image into the MSB and LSB branches and deals with them separately with uniform quantization until the final addition, which prevents each branch from getting information from the other. MuLUT~\cite{li2022mulut} utilizes quantization and interpolation to build the connection between stages, which requires later fine-tuning to compensate for the loss of information caused by quantization.}

Here, we propose a new multistage framework as illustrated in Fig.~\ref{fig:overall_architecture}. In each stage, we split the 8-bit input image into MSB and LSB branches for different LUT blocks to extract features and upscale the resolution. The features from both branches are then combined to generate the complete enlarged image. The input and output of each stage are 8-bit images with different resolutions, unless the upscaling factor is 1. This module can be stacked to create cascaded stages, where the latter stage can access information from the two branches of the former stages. Residual connections can be conveniently implemented. The entire flow does not require interpolation, which significantly speeds up inference.

Furthermore, previous works utilize cascaded stages to increase the RF for a better performance, where upsampling is usually done in the last step. It is intuitively difficult to reproduce a high-resolution patch from a few pixels in one single step, especially when the upscaling factor is large. Besides, the storage space of LUT is proportional to the square of the upscaling factor $r$, we can save half the storage space by dividing the upscaling factor $4\times$ (viz. $4\times4=16$) into two factors $2\times$ (viz. $2\times2+2\times2=8$). This simple yet previously unexplored progressive upsampling trick gives another booster to our proposed SISR scheme. Table~\ref{tab:progressive_upsample} verifies the effectiveness of progressive upsampling in our LUT-based methods, along with further savings in storage.

\begin{table*}[th]
\centering
\resizebox{1.9\columnwidth}{!}{%
\begin{tabular}{|cc|ccc|ccccc|}
\hline
\multicolumn{2}{|c|}{Method} & RF  & \# Pixels & \multirow{2}{*}{Storage} & Set 5 & Set 14 & BSDS100 & Urban100 & Manga109\\
MSB & LSB & MSB/LSB &  MSB/LSB & ~ & PSNR/SSIM & PSNR/SSIM  & PSNR/SSIM  & PSNR/SSIM  & PSNR/SSIM \\
\hline
HD & HD & $3\times3$/$3\times3$    & 2/2 & 16KB & 29.39/0.835 & 26.79/0.730 & 26.38/0.692 & 23.73/0.689 &  26.28/0.825 \\
HD & HDB & $3\times3$/$5\times5$    & 2/3 & 200KB &  29.40/0.835 & 26.79/0.730 & 26.39/0.692 & 23.73/0.689 &  26.29/0.825 \\
HDB & HD & $5\times5$/$3\times3$    & 3/2 & 200KB & 29.98/0.848	& 27.13/0.741 & 26.61/0.702 & 24.01/0.700 & 26.88/0.837 \\
HDB & HDB  & $5\times5$/$5\times5$    & 3/3 & 384KB  & 29.99/0.849 & 27.15/0.742 & 26.63/0.703 & 24.02/0.701 & 26.90/0.839 \\
\hline
\end{tabular}%
}
\vspace{2mm}
\caption{Asymmetric parallel structure: a smaller LSB-branch RF yields negligible performance drops but large storage savings.}
\label{tab:effect_of_asymmetric_parallel}
\vspace{3mm}
\centering
\resizebox{1.91\columnwidth}{!}{%
\begin{tabular}{|cc|cc|ccccc|}
\hline
\multirow{2}{*}{\# Stages} & \multirow{2}{*}{Upscale} & \multirow{2}{*}{Storage}& \multirow{2}{*}{Runtime} & Set5 & Set14  & BSDS100 & Urban100 & Manga109 \\
& ~ & ~ & ~  & PSNR/SSIM & PSNR/SSIM & PSNR/SSIM & PSNR/SSIM & PSNR/SSIM \\
\hline
$1$ & $4$ & 200KB & 108.8ms &  29.97/0.84 & 27.13/0.741 & 26.61/0.702 & 24.01/0.700 & 26.86/0.837 \\
$2$ & $1\times4$ & 213KB & 115.5ms & 29.99/0.848 & 27.14/0.742 & 26.62/0.702 & 24.02/0.700 & 26.86/0.837 \\
\rowcolor{lime}
 $2$ & $2\times2$ & 100KB & 201.7ms & 30.35/0.859 & 27.39/0.748 & 26.73/0.706 & 24.23/0.711 & 27.38/0.852  \\
\hline
 $3$ & $1\times2\times2$ & 112.5KB & 212.1ms & 30.34/0.860 & 27.40/0.748 & 26.75/0.707& 24.25/0.712 & 27.46/0.853 \\
\rowcolor{LightCyan}
\rowcolor{LightCyan}
 $3$ & $2\times1\times2$ & 112.5KB & 235.3ms & 30.41/0.860 & 27.44/0.749 & 26.78/0.707& 24.27/0.713 & 27.51/0.854 \\
 $3$ & $2\times2\times1$ & 112.5KB & 475.0ms & 30.43/0.861 & 27.47/0.750 & 26.79/0.707 & 24.31/0.713 & 27.58/0.856 \\
\hline
\end{tabular}%
}
\vspace{1mm}
\caption{Ablating various progressive upsampling strategies. Runtime is measured  for outputting a $1280\times720 $ image on a desktop CPU averaged across 10 runs. We refer to models in the \colorbox{lime}{lime} and \colorbox{LightCyan}{blue} rows as HKLUT-S and HKLUT-L, respectively.}
\label{tab:progressive_upsample}
\end{table*}

\begin{table*}[t]
\centering
\resizebox{1.95\columnwidth}{!}{%
\begin{tabular}{|l|cc|ccccc|}
\hline
\multirow{2}{*}{Method}  & \multirow{2}{*}{Size} & \multirow{2}{*}{Ratio} & Set5 & Set14  & BSDS100 & Urban100 & Manga109 \\
~ &~ &~ &  PSNR/SSIM & PSNR/SSIM & PSNR/SSIM & PSNR/SSIM & PSNR/SSIM \\
\hline\hline
Nearest~\cite{bevilacqua2012low}  & - & - & 26.25/0.737 & 24.65/0.653 & 25.03/0.629 & 22.17/0.615 & 23.45/0.741 \\
 Bilinear~\cite{kirkland2010bilinear}  & - &  - & 27.55/0.788 & 25.42/0.679 & 25.54/0.646 & 22.69/0.635 & 24.21/0.767 \\
 Bicubic~\cite{keys1981cubic}  & - &  - & 28.42/0.810 & 26.00/0.702 & 25.96/0.667 & 23.14/0.657 & 24.91/0.787\\
\hline\hline
 ANR~\cite{timofte2013anchored}  & 1.43MB & 14.64 & 29.70/0.842 & 26.86/0.737 & 26.52/0.699 & 23.89/0.696 & 26.18/0.821 \\
 A+~\cite{timofte2015a+}  & 15.17MB& 155.34 & 30.27/0.860 & 27.30/0.750 & 26.73/0.709 & 24.33/0.719 & 26.91/0.848 \\
\hline\hline
 CARN-M~\cite{ahn2018fast} & 1.59MB & 16.28 & 31.82/0.890 & 28.29/0.775 & 27.42/0.731 & 25.62/0.769 & 29.85/0.899\\
 RRDB~\cite{wang2018esrgan} & 63.83MB & 653.62 & 32.60/0.900 & 28.88/0.790 & 27.76/0.743 & 26.73/0.807 & 31.16/0.916 \\
\hline\hline
SRLUT~\cite{jo2021srlut}  & 1.27MB & 13.05 & 29.82/0.847 & 27.01/0.736 & 26.53/0.695 & 24.02/0.699 & 26.80/0.838 \\
SPLUT-S~\cite{ma2022splut}  & 5.5MB & 56.32 &  30.01/0.852  & 27.20/0.743 & 26.68/0.702 & 24.13/0.706 & 27.00 0.843\\
SPLUT-M~\cite{ma2022splut}  & 7MB &  71.68 &  30.23/0.857 & 27.32/0.746 & 26.74/0.704 & 24.21/0.709 & 27.20 0.848 \\
SPLUT-L~\cite{ma2022splut}  & 18MB & 184.32 &  30.52/0.863 & 27.54.0.752 & 26.87/0.709 & 24.46/0.719 & 27.70 0.858 \\
MuLUT-SDY~\cite{li2022mulut}  & 3.82MB & 39.15 &  30.40/0.860 & 27.48/0.751 & 26.79/0.709 & 24.31/0.714 & 27.52/0.855 \\
MuLUT-SDY-X2~\cite{li2022mulut}  &  4.06MB & 41.60 &  30.60/0.865 & 27.60/0.754 & 26.86/0.711 & 24.46/0.719 & 27.90/0.863 \\
RCLUT~\cite{liu2023rclut}  &  1.51MB & 15.46 &  30.72/0.868 & 27.67/0.758 & 26.95/0.715 & 24.57/0.725 & 28.05/0.8655 \\
HKLUT-S (\textbf{ours}) &  \textbf{100 KB} & 1 & 30.35/0.859 & 27.39/0.748 & 26.73/0.706 & 24.23/0.711 & 27.38/0.852  \\
HKLUT-L (\textbf{ours}) &  \textbf{112.5 KB} & 1.13 &  30.41/0.860 & 27.44/0.749 & 26.78/0.707& 24.27/0.713 & 27.51/0.854  \\
\hline
\end{tabular}%
}
\vspace{1mm}
\caption{Different SR methods on 5 benchmark datasets. The size ratio is w.r.t. HKLUT-S. The 4 groups from top to bottom are interpolation, sparse-coding, CNN \& LUT approaches, respectively.}
\label{tab:final_results}
\end{table*}

\section{Experiments}
\label{sec:exp}

\subsection{Implementation Details}
\noindent{\bf Datasets and Metrics.} We train the proposed HKLUT on the DIV2K dataset~\cite{agustsson2017ntire}, a popular dataset in the SR field. The DIV2K dataset provides 800 training images and 100 validation images with 2K resolution and the corresponding downsampled images. We use the widely adopted Peak Signal-to-Noise Ratio (PSNR) and structural similarity index (SSIM)~\cite{wang2004image} as evaluation metrics. Five well-known datasets: Set5~\cite{bevilacqua2012low}, Set14~\cite{zeyde2012single}, BSDS100~\cite{martin2001database}, Urban100~\cite{huang2015single} and Manga109~\cite{matsui2017sketch} are benchmarked. Following other LUT-based methods, we mainly focus on SR tasks with an upscaling factor of 4. We also report the runtime, theoretical energy cost, and peak memory of generating a $1280\times720$ HR image from a $640\times360$ LR image.

\noindent{\bf Experimental setting.} Before converting into LUTs, the network is trained with 800 training images in DIV2K for 200k iterations with a batch size of 16 on Nvidia RTX 3090 GPUs. We utilize the Adam optimizer~\cite{kingma2014adam} ($\beta_1 = 0.9$, $\beta_2 = 0.999$ and $\epsilon = 1e-8$) with the MSE loss to train the HKLUT. The initial learning rate is set to $5\times10^{-4}$, which decays to one-tenth after 100k and 150k iterations, respectively. We randomly crop LR images into $48\times48$ patches as input and enhance the dataset by random rotation and flipping. The runtime is measured on an Intel Core i5-10505 CPU with 16GB RAM, averaged over 10 runs.

\noindent{\bf Baselines.} Aligning with prior work, we evaluate HKLUT against several well-recognized SISR methods, such as interpolation-based methods (nearest neighbor~\cite{bevilacqua2012low}, bilinear~\cite{kirkland2010bilinear} and bicubic~\cite{keys1981cubic} interpolation), sparse-coding-based methods (ANR~\cite{timofte2013anchored} and A+~\cite{timofte2015a+}), DNN-based methods (CARN~\cite{du2022fast} and RRDB~\cite{wang2018esrgan}), and LUT-based methods (SRLUT~\cite{jo2021srlut}, SPLUT~\cite{ma2022splut}, MuLUT~\cite{li2022mulut} and RCLUT~\cite{liu2023rclut}).

\subsection{Quantitative Results}
\label{subsec:quantitative_results}
{\bf Asymmetric Parallel Structure}. To demonstrate the effectiveness of our proposed asymmetric parallel structure, we utilize the 2-pixel HDLUT with $3\times3$ RF and 3-pixel HDBLUT with $5\times5$ RF to investigate the relationship between RF, LSB, and MSB for the $4\times$ SR. Table~\ref{tab:effect_of_asymmetric_parallel} shows that using HDLUTs for both branches results in worse performance (29.39dB for Set5). A negligible increase (+0.01dB) in performance is achieved when we replace the LSB branch with the larger-RF HDBLUT, at a cost of over 12 times the growth in model size (200KB vs. 16KB). However, by simply swapping the two branches, i.e., using HDBLUT for MSB and HDLUT for LSB, we obtain a significant boost in performance (+0.58dB) with the same storage and computing. When using HDBLUT for both MSB and LSB branches, we only obtain a marginal increase in PSNR (+0.01dB), at a cost of nearly doubling the storage (384KB vs. 200KB). This validates our conjecture that a 2-pixel LUT is sufficient for the LSB because of the small ERF, while a 3-pixel LUT is necessary for the MSB to capture contextual information such as texture. In short, the proposed asymmetric parallel structure allows for the efficient allocation of resources and saves half of the storage space without sacrificing performance.

\noindent{\bf Multistage.} \textcolor{black}{Table~\ref{tab:progressive_upsample} examines various upsampling strategies to demonstrate the effectiveness of progressive upsampling. Following the asymmetric parallel structure, HDBLUT and HDLUT are used for each stage's MSB and LSB branches, respectively. Comparing the 1st and 3rd rows, using an asymmetric parallel structure, 2-stage progressive upsampling ($2 \times 2$) shows a clear advantage in terms of SR performance compared to 1-stage upsampling ($4\times$) (30.35dB vs. 29.97dB on Set 5), while requiring only half the storage space (100KB vs. 200KB). The advantage of progressive upsampling becomes even clearer when compared to results on the 2nd row, which corresponds to the same 2-stage but with upsampling in a single step ($1\times4$) (-0.36dB). 
Our results show that $2\times1\times2$ upsampling achieves the best tradeoff by considering storage, runtime, and performance. We, therefore, refer to models listed on the 3rd and 5th rows as HKLUT-S (highlighted in lime) and HKLUT-L (highlighted in blue), respectively.}

\noindent{\bf Performance and Storage} In Table~\ref{tab:final_results}, we compared our proposed HKLUT with well-recognized SISR approaches such as interpolation, sparse-coding, and CNNs, as well as the latest LUT-based methods. Our approach has a clear advantage in storage, which aligns with the objective of this paper. HKLUT-S, being the smallest parameterized model, only requires 100KB, which is $10\times$ less than the second smallest SRLUT (1.27MB) with a much better performance. Both HKLUT-S and HKLUT-L outperform all classical baselines, including A+, which is significantly larger than HKLUT-S (15.17MB vs. 100KB). Although CNN-based approaches exhibit good performance in terms of PSNR and SSIM, they often require larger storage. For instance, RRDB is $650\times$ larger than our HKLUT-S, making it challenging to deploy on edge devices. Compared to LUT-based approaches, our methods not only require less storage but also achieve comparable performance. HKLUT-L outperforms MuLUT-SDY on the Set5 dataset (30.41dB vs. 30.40dB) with $35\times$ less storage. Despite being $70\times$ smaller, HKLUT-S significantly outperforms SPLUT-M (30.35dB vs. 30.23dB).

\begin{figure}[t]
    \centering
    \includegraphics[width=.8\columnwidth]{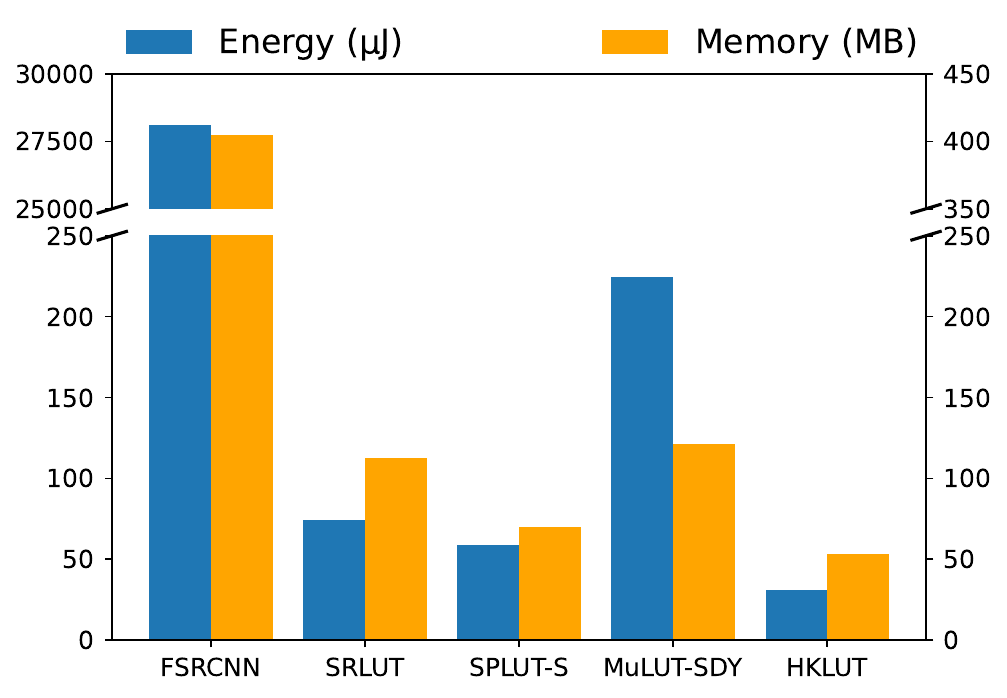}
    \caption[The LOF caption]{Energy cost and peak memory to output a $1280\times720$ image through $2\times$ SR.}\vspace{-12pt}
    \label{fig:energy_memory}
\end{figure}

\noindent{\bf Energy \& Peak Memory.} Following the setting in MuLUT~\cite{li2022mulut}, Fig.~\ref{fig:energy_memory} shows the theoretical energy costs and peak memory usages of FSRCNN (32.70dB) and several LUT-based methods (31.73dB$\sim$32.35dB) for generating a $1280\times720$ image through $2\times$ SR. We estimated the theoretical energy cost using the same protocol as AdderSR~\cite{song2021addersr}, and computed the peak memory usage using \textit{memray}\footnotemark[1]. For LUT-based approaches, we used their smallest variants. One of the major advantages of LUT-based methods is their ability to drastically reduce energy and memory consumption. Even FSRCNN, the simplest CNN for SISR, consumes $900\times$ and $7.5\times$ more energy and memory compared to HKLUT. 
The feature map aggregation module of SPLUT not only instantiates larger LUTs but also requires floating-point operations, resulting in increased energy consumption and peak memory usage. Meanwhile, both SRLUT and MuLUT consume a significant amount of energy for the interpolation operation. In contrast, HKLUT gets rid of energy-intensive interpolation and solely relies on integer operations. Fig.~\ref{fig:energy_memory} shows that HKLUT exhibits the lowest energy cost and peak memory usage amongst all competitors. 
\footnotetext[1]{Measured with \textit{memray} (\url{https://github.com/bloomberg/memray}) on a desktop CPU.}

\begin{figure*}[t]
    \centering
    \includegraphics[width=1.9\columnwidth]{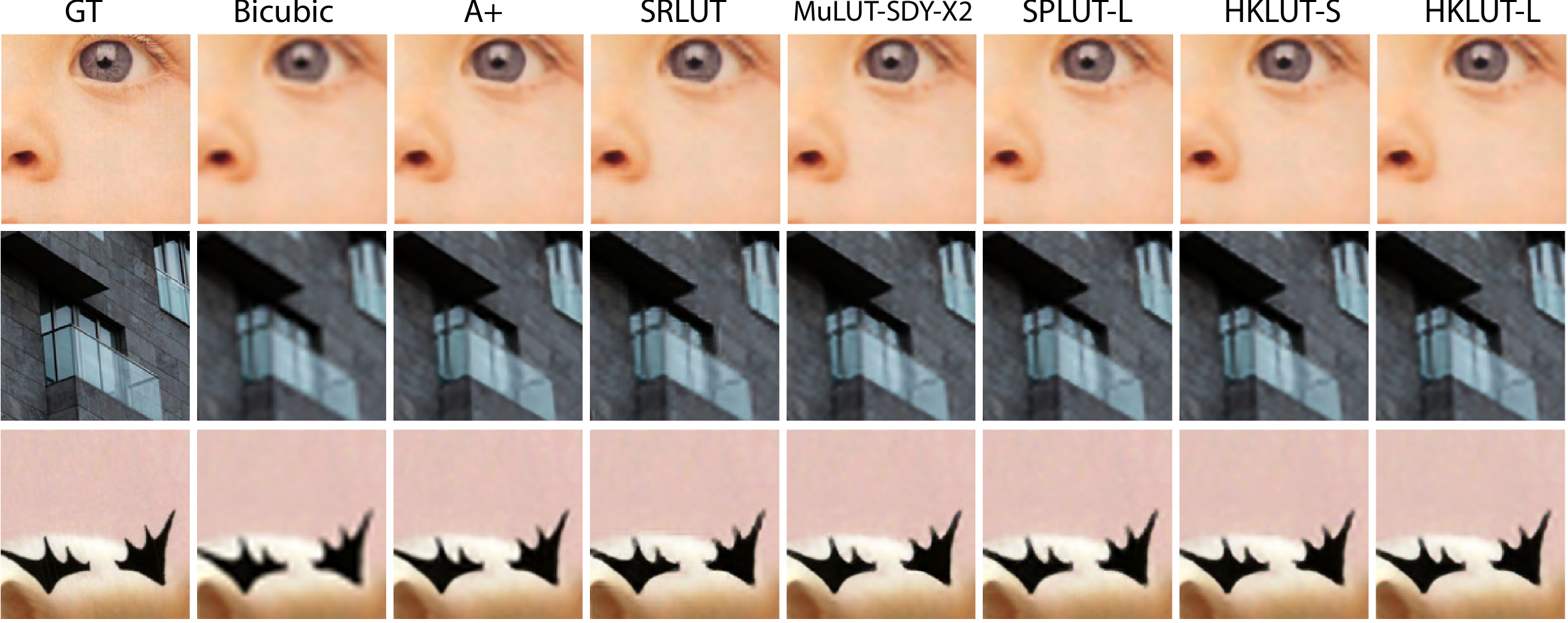}
    \caption{Qualitative results on selected images from Set5, Urban100 and Manga100 datasets.}
    \label{fig:qualitative_results}
\end{figure*}

\noindent{\bf Runtime.} Using the same settings as for calculating energy and peak memory, we compared the runtimes of FSRCNN and LUT-based approaches on both a desktop CPU and the Raspberry Pi 4 Model B. To ensure a fair comparison, we used the official Python code released by the authors, without relying on native mobile implementation or any speedup tricks. We recorded the average runtime of generating a $1280 \times 720$ image from a $640\times 360$ image in Table~\ref{tab:runtime}. It is clear that the 4-simplex interpolation bottlenecks the inference speed of SRLUT and MuLUT at the final stage. SPLUT and our approaches, featuring parallel branches, exhibit shorter inference times on Raspberry Pi and CPU. Unsurprisingly, HKLUT achieves the fastest inference speed.

\begin{table}[t]
    \centering
    \resizebox{1\columnwidth}{!}{%
    \begin{tabular}{|l|ccccc|}
    \hline
    Method &  FSRCNN & SRLUT & SPLUT-S & MuLUT-SDY & HKLUT \\
    \hline
    CPU (ms) & 332.64 & 1412.24 & 197.49 & 5128.17 & \textbf{148.67} \\ 
    Raspberry (ms) & 3589.86 & 8021.56 & 1096.73 & 35637.12 & \textbf{908.52} \\
    
    \hline
    \end{tabular}%
    }
    \caption{Comparing runtimes for generating a $1280 \times 720$ image through $2 \times$ SR. Results are obtained by averaging across 10 runs.}
    \label{tab:runtime}
\end{table}


\subsection{Qualitative Results}
Fig.~\ref{fig:qualitative_results} presents a visual comparison between the bicubic interpolation baseline and different LUT-based approaches on images selected from five benchmark datasets. The results clearly demonstrate that LUT-based approaches improve sharpness compared to bicubic interpolation which generates blurry images. The 3rd row (Manga109) shows that the SRLUT approach exhibits severe rippling artifacts around the edges, which may be caused by its limited RF. HKLUT-S produces similar sharpness to MuLUT-SDY-X2 and SPLUT-L, despite being more than $40\times$ and $180\times$ smaller, respectively. 

\section{Ablation Studies}
\label{sec:ablation}

\subsection{Ablation on Kernel Shapes} 
\label{subsec:ablation_kernel_shape}
\begin{figure}[th]
    \centering
    \includegraphics[width=0.85\columnwidth]{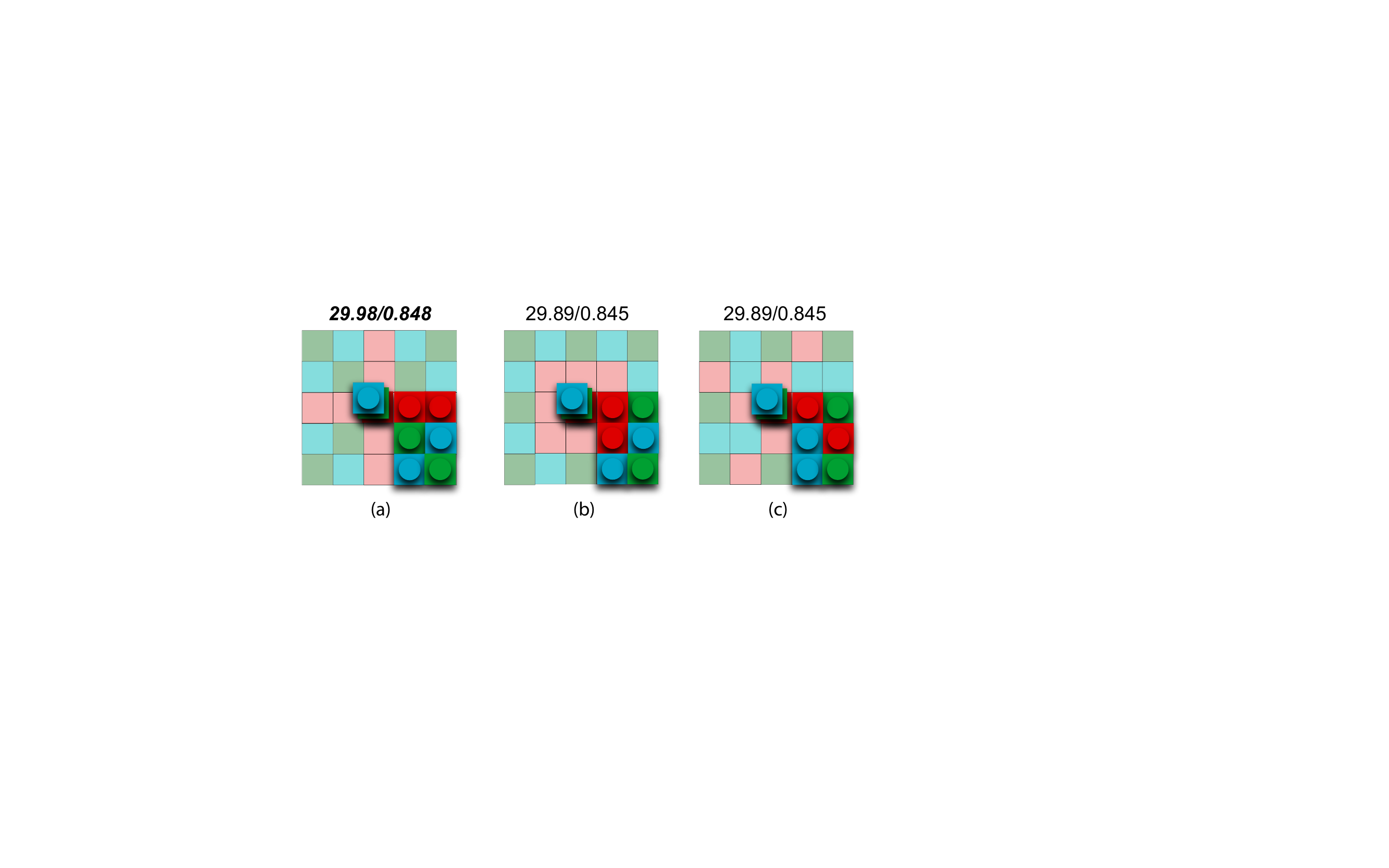}
    \caption{The 3 best-performing kernel designs for MSB are reported, along with the PSNR/SSIM for $4\times$ SR on the Set5 dataset.}
    \label{fig:hklut_abc}
\end{figure}

To obtain the HDB kernel for our MSB branch, we conducted an ablation study on kernel shape. Instead of exhausting all possible combinations, we employed softmax during training with an annealing temperature to select the most important pixel per column. After training, we evaluated the binarized kernels. We performed 20 runs with different random seeds using a one-stage model, with the LSB branch fixed as the HD kernel. Each run consisted of 20k (10\%) training iterations. Next, we selected the top 3 best-performing candidates with non-overlapping kernels and retrained them from scratch for 200k iterations. Fig.~\ref{fig:hklut_abc} shows the 3 best-performing for MSB branch, where HDB (i.e. (a)) ranks as the best.

\begin{table}[th]
\centering
\resizebox{0.95\linewidth}{!}{
\begin{tabular}{|c|ccc|}
\hline
\multirow{2}{*}{Cascading Approach}  &  Set5 & Set14  & BSDS100 \\
 ~ & PSNR/SSIM & PSNR/SSIM  & PSNR/SSIM \\
\hline\hline
HKLUT & \textbf{30.35/0.859} & \textbf{27.39/0.748} & \textbf{26.73/0.706} \\
SPLUT & 29.32/0.836 & 26.55/0.724 & 26.15/0.686\\
\hline
\end{tabular}%
}
\caption{Ablation on different ways of cascading LUTs. HDBLUT and HDLUT are used for MSB and LSB branch, respectively.} 
\label{tab:ablation_cascade_lut}
\end{table}

\subsection{Ablation on LUT Cascading} 
In previous work~\cite{ma2022splut}, a series-parallel structure was proposed to remove the need for interpolation. This structure divides an 8-bit image into MSB and LSB parts and processes them separately. The series structure involves cascading LUTs within each branch. Communication between branches only occurs at the end of the model to generate the final SR image by combining information. The output activations of each LUT in intermediate stages are clipped between [-8, 7] and quantized into 16 levels. However, we believe that this approach is counter-intuitive and suboptimal. On the other hand, our method allows communication between branches during stage transition. The output of each stage is clipped between [0, 255] represented in INT8, enabling meaningful results even in intermediate stages. Table~\ref{tab:ablation_cascade_lut} demonstrates the superior performance of our approach compared to the multistage structure in SPLUT~\cite{ma2022splut} (30.35dB vs. 29.32dB for Set5), both using the $2\times2$ progressive upsampling technique. 


\section{Conclusion}
This paper aims to develop minimalist LUTs for efficient LUT-based SISR. We propose a new class of models called HKLUTs employing kernels with fewer input pixels while retaining the RF, resulting in an exponential reduction in storage without compromising performance. In particular, HKLUT consists of an asymmetric parallel structure that employs big-little kernel patterns to adapt to the underlying characteristics of MSB and LSB branches. A progressive multistage architecture, without expensive interpolation between stages, is devised to reduce storage by half. By seamlessly fusing these innovations, HKLUT models set the first record to attain $>30$dB on the Set 5 dataset at merely one-tenth of 1MB storage. Extensive experiments verify that HKLUT is a versatile alternative to previous large LUT-based approaches on resource-constrained devices.

\newpage






\bibliographystyle{named}
\bibliography{ijcai24}

\clearpage
\appendix

\centerline{\textbf{\scalebox{1.5}{Appendix}}}

\section{Details of Neural Nets during Training}
\label{appendix:training_details}
\begin{figure}[th]
    \centering
    \includegraphics[width=0.9\columnwidth]{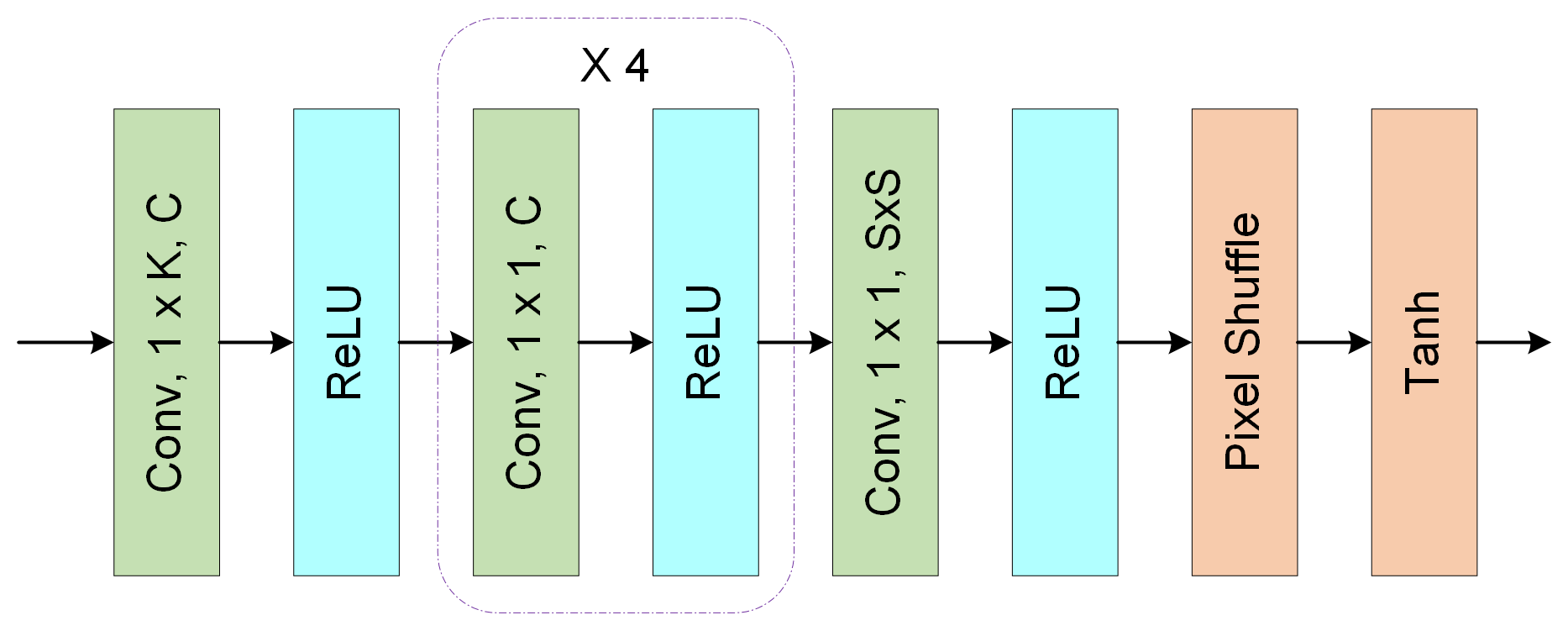}
    \caption{Detailed architecture of the neural network employed during training.}
    \label{fig:lut-net}
\end{figure}

During training, each look-up table (LUT) is replaced with a simple convolutional neural network (CNN), as shown in Fig.~\ref{fig:lut-net}. The CNN consists of one convolution (CONV) layer with a $1\times K$ kernel, and five $1\times 1$ CONV layers, interleaved with RELU~\cite{glorot2011deep} nonlinearity. No dense and residual connections are used here. $K$, $C$, and $S$ represent the number of input pixels to the LUT, the output channel number of the CONV layer (i.e., 64), and the upscaling factor, respectively. Since the LUT only accepts a finite number (K) of input pixels, all convolution kernels are of size $1\times 1$, except for the first one, which is of size $1\times K$. If the kernel shape of the first convolution is neither horizontal nor vertical, the pixels are first sampled according to the defined index pattern and then reshaped to $1\times K$. After the last CONV layer, pixel shuffle~\cite{shi2016shuffle} is applied to generate the high-resolution (HR) image. A Tanh activation function is used at the end to project the output to a range between -1 and 1. To cache CNN outputs into a LUT, they are converted from FP32 to INT8. This is done by multiplying the FP32 values by 127 to ensure they fit within the INT8 range (-128 to 127). The resulting numbers are rounded down to the nearest integer and stored in the LUT as INT8.

\begin{figure}[ht]
    \centering
    \subfigure[Set14]{
        \includegraphics[width=0.47\columnwidth]{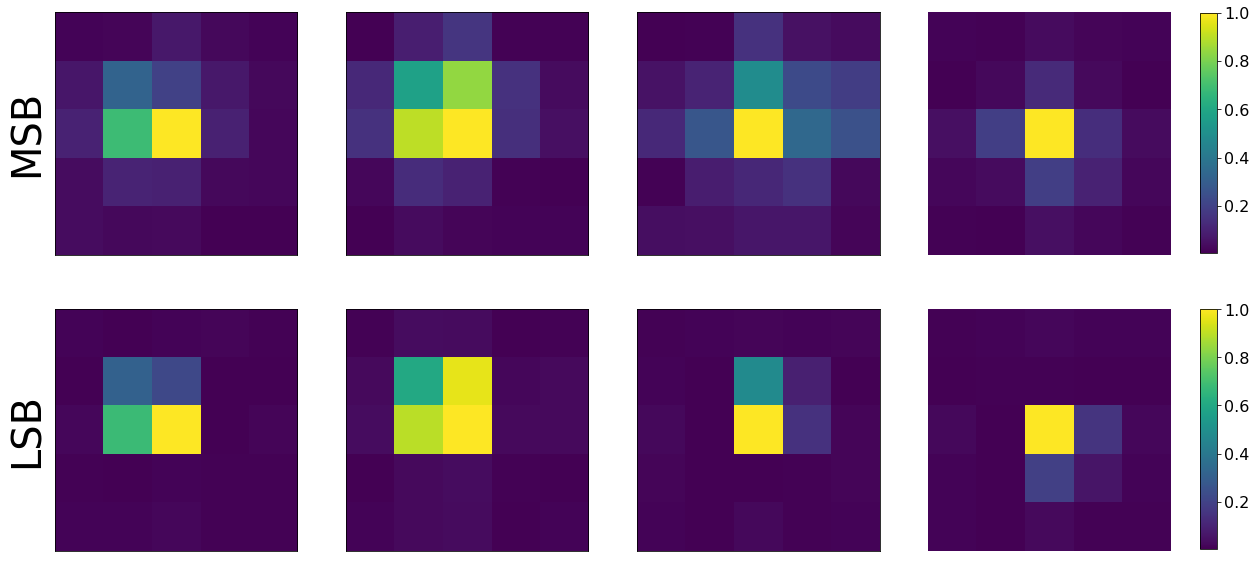}
    }
    \subfigure[B100]{
	\includegraphics[width=0.47\columnwidth]{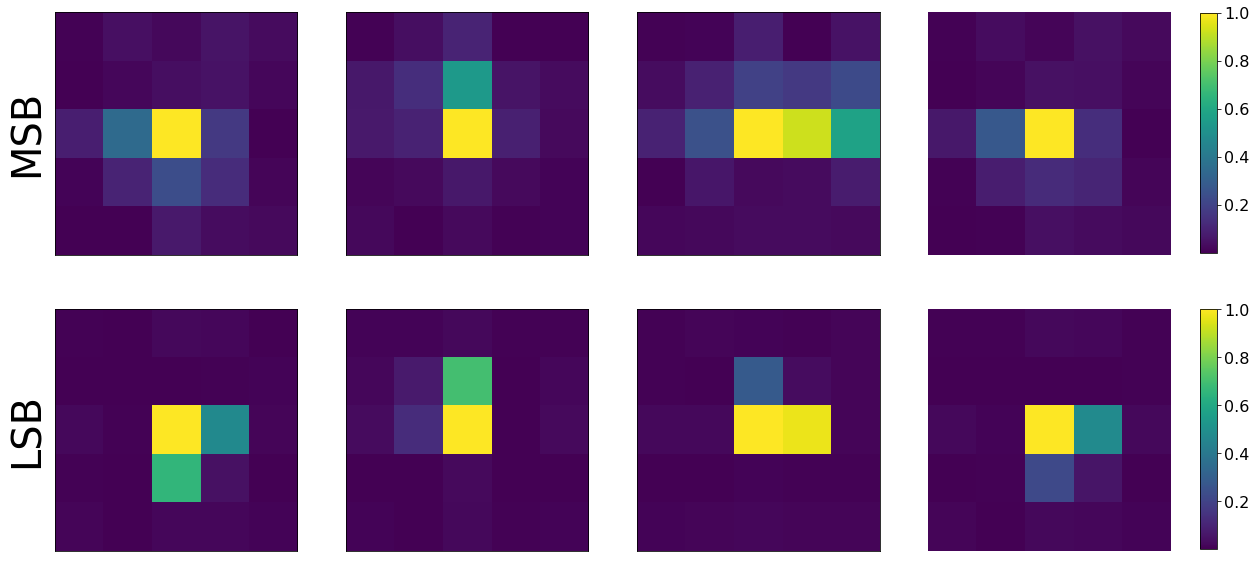}
    }
    \quad    
    \subfigure[Urban100]{
        \includegraphics[width=0.47\columnwidth]{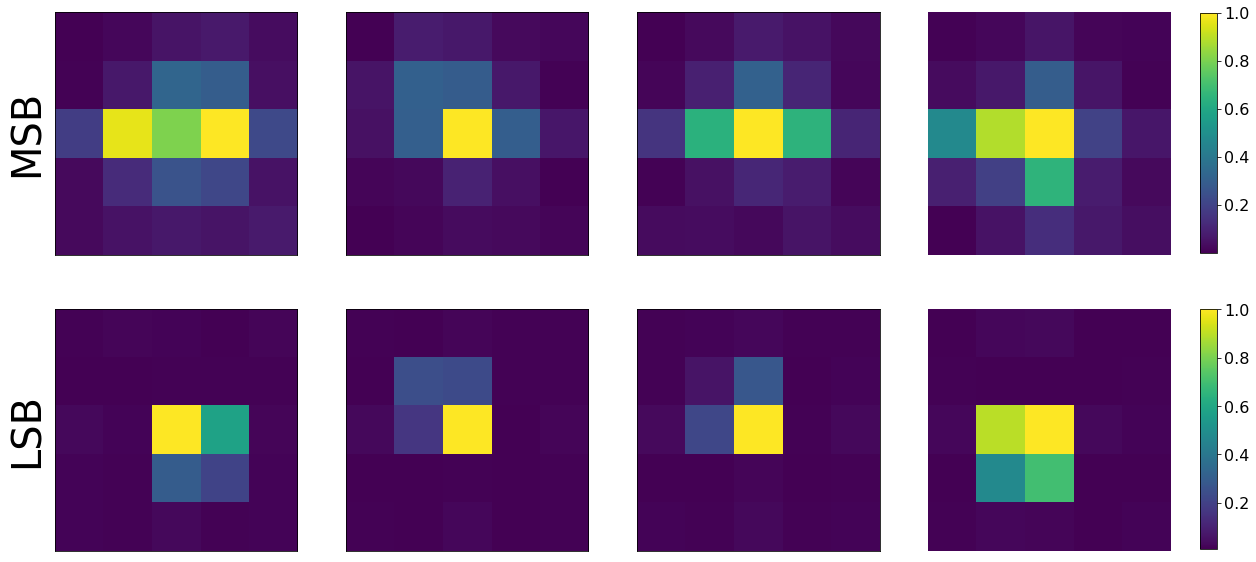}
    }
    \subfigure[Manga109]{
	\includegraphics[width=0.47\columnwidth]{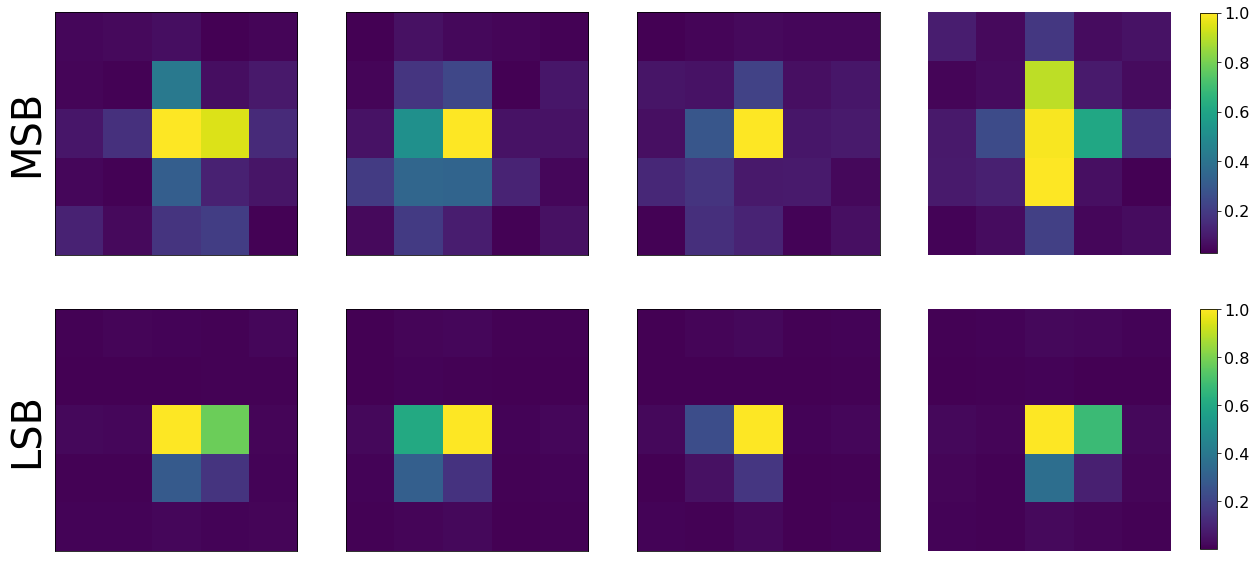}
    }
    \caption{Four randomly selected ERFs obtained from images in the Set14, B100, Urban100, and Manga109 datasets. \textbf{Top}: MSB. \textbf{Bottom}: LSB. The brightness denotes the model’s sensitivity to that pixel.}
    \label{fig:erf}
\end{figure}

\section{Effective Receptive Field}
As elaborated in the manuscript, SPLUT~\cite{ma2022splut} neglects the disparity between the information from the most-significant-bit (MSB) and the least-significant-bit (LSB) branches, despite their identical theoretical receptive fields (e.g., $5\times5$). We hypothesize that using the same set of LUTs for both MSB and LSB branches may not be the optimal choice. To find an ideal solution, we visualize the effective receptive fields (ERFs) of both branches and introduce an asymmetric structure based on this observation, aiming to substantially reduce the size of LUTs. In Fig.~\ref{fig:erf}, we provide additional demonstrations of the ERFs of various images across datasets, confirming that the MSB branch has a larger ERF while the LSB branch maintains a more compact one.

\vspace{-2mm}

\section{LUT Design Comparison}
\begin{figure}[ht]
    \centering
    \subfigure[SRLUT]{
        \includegraphics[width=0.42\columnwidth]{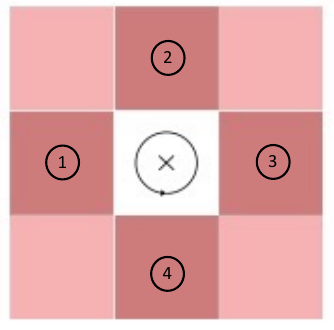}
    }
    \subfigure[HDLUT]{
	\includegraphics[width=0.42\columnwidth]{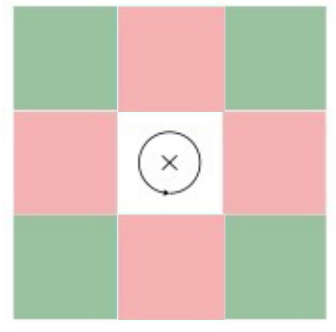}
    }
    \quad    
    \subfigure[MuLUT]{
        \includegraphics[width=0.42\columnwidth]{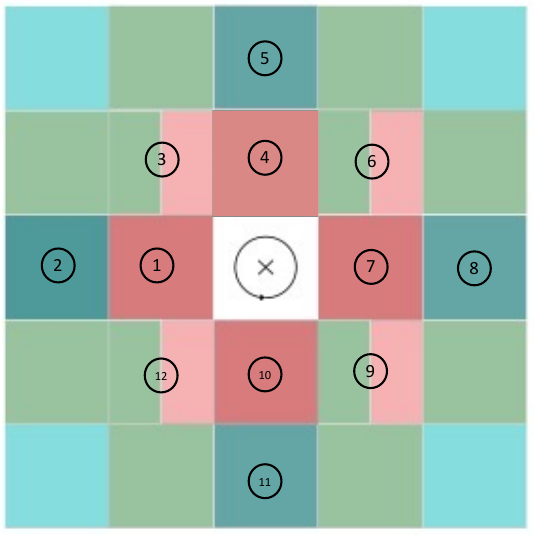}
    }
    \subfigure[HDBLUT]{
	\includegraphics[width=0.42\columnwidth]{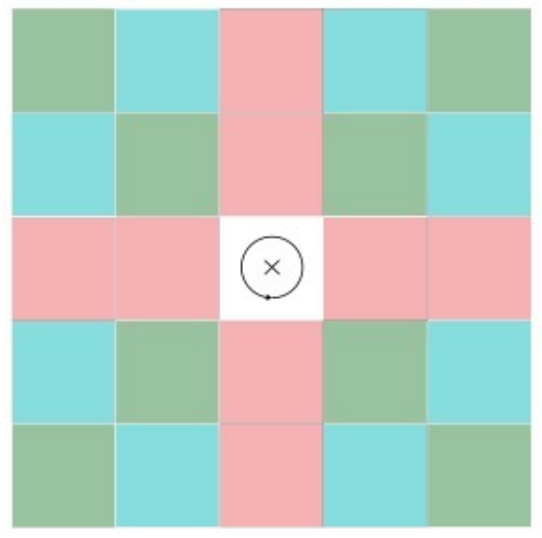}
    }
    \caption{A comparison of various LUTs' RF patterns. The numbers indicate all overlapping pixels, with darker pixels showing locations where the same kernel overlaps with itself during rotation ensemble, while pixels with two distinct colors denote areas where different kernels overlap with each other. In each case, the center pixel is used as a pivot for rotation by all kernels.}
    \label{fig:rf_comparison}
\end{figure}

In this section, we highlight the differences in LUT designs used in various LUT-based SISR methods. Fig.~\ref{fig:rf_comparison} shows a comparison of different LUT designs. Unlike SRLUT and MuLUT, which use a $3\times3$ RF with 4 overlapping pixels and a $5\times5$ RF with 12 overlapping pixels, respectively, our designs, HDLUT and HDBLUT, aim to match the same RF with minimal input pixels. This achieves a near-optimal tradeoff between memory footprint and performance.

\section{Architectural Design Comparison}

\begin{figure*}[t!]
    \centering
    \includegraphics[width=1.6\columnwidth]{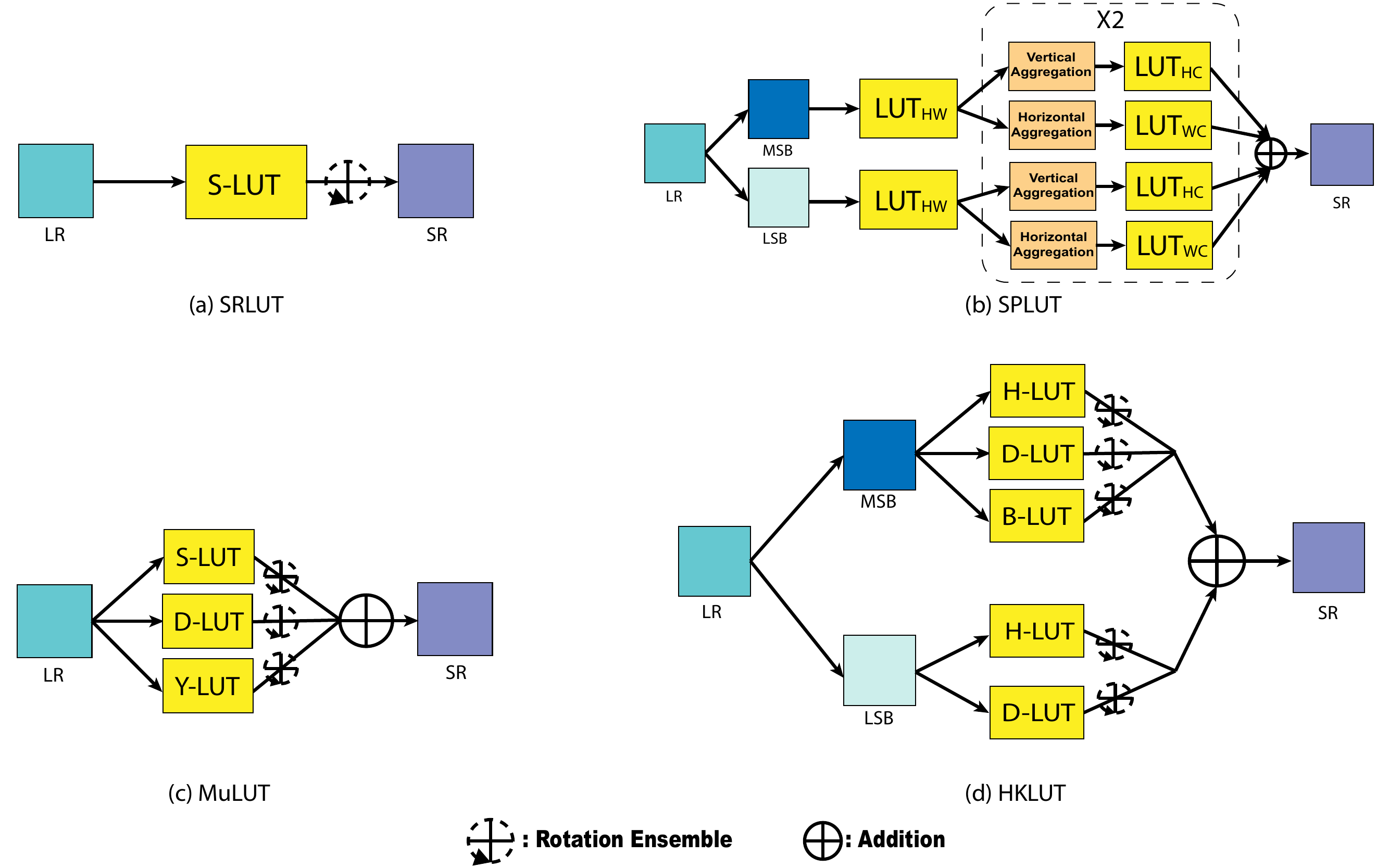}
    \caption{Comparison of architectural designs. Residual connections are omitted for brevity. (a) SRLUT (b) SPLUT (c) MuLUT (d) HKLUT (Ours)}
    \label{fig:architecture_comparison}
\end{figure*}

This section explores various LUT-based SISR architectures. Fig.~\ref{fig:architecture_comparison} depicts these architectures. LUT here refers to a CNN similar to that in Fig.~\ref{fig:lut-net} with the prefix or subscript describing the kernel shape of the first CONV layer. Fig.~\ref{fig:architecture_comparison} (a) depicts the overall structure of SRLUT~\cite{jo2021srlut}, which consists of only a single LUT, corresponding to a CNN with the first layer using a $2\times2$ square kernel. During training and inference, rotation ensemble is used to expand the receptive field (RF) from $2\times2$ to $3\times3$. MuLUT~\cite{li2022mulut}, as shown in Fig.~\ref{fig:architecture_comparison} (c), extends SRLUT by using multiple LUTs with different input kernel shapes. These kernels have complementary indexing patterns, as they can fill up the entire $5\times5$ area using the rotation ensemble trick. The same structure can be stacked sequentially to form a multistage structure to further enlarge the RF. Fig.~\ref{fig:architecture_comparison} (b) demonstrates SPLUT~\cite{ma2022splut} which divides an 8-bit image into two branches: one with the most significant 4 bits (MSB) and the other with the least significant 4 bits (LSB) to eliminate interpolation at the final stage. To maintain consistency, we follow the terminology from SPLUT~\cite{ma2022splut}. Here, $\rm{LUT}_{HW}$ is referred to as spatial lookup, which uses the $2\times2$ square kernel as the S-LUT in Fig.~\ref{fig:architecture_comparison} (a) and (c), but with multi-channel outputs. The number of output channels can be either 4 (SPLUT-S), 8 (SPLUT-M), or 16 (SPLUT-L). 
The multi-channel feature maps are then grouped along the channel dimension and fed to either vertical or horizontal aggregation. The aggregated features are then fed to $\rm{LUT}_{HC}$ and $\rm{LUT}_{WC}$, both of which have a $2\times2$ input pattern. The input pattern of $\rm{LUT}_{HC}$ is along both the height and channel dimensions, while that of $\rm{LUT}_{WC}$ is along both the width and channel dimensions. The aggregation modules, $\rm{LUT}_{WC}$ and $\rm{LUT}_{HC}$, can be repeated multiple times in series to further expand the RF. The final super-resolved (SR) image is produced by summing the outputs of the two branches. The proposed HKLUT architecture is illustrated in Fig.~\ref{fig:architecture_comparison} (d). Our method utilizes the two-branch structure of SPLUT to eliminate the interpolation step. However, instead of employing the same set of LUTs in both branches, we propose an asymmetric parallel structure that uses smaller LUTs for the LSB branch to reduce storage space. Within each branch, we use multiple LUTs with the rotation ensemble trick similar to MuLUT. However, LUTs of our design have fewer input pixels, resulting in an exponential reduction in size. Summing the results from the two branches yields the final SR image. The entire structure is treated as a building block. A multistage structure is constructed by repeating the whole module. This allows convenient implementation of different progressive upsample strategies to further reduce storage.

\section{Lightweight CNN Comparisons}
\label{appendix:lightweight_cnns}
In Section~\ref{sec:exp}, we selected RRDB (best accuracy) and CARN-M (storage-accuracy balance) to align with previous literature (SRLUT, SPLUT, and MuLUT) using the same set of baselines. We listed the storage, PSNR (on BSDS100), energy cost, and peak memory of two state-of-the-art (SOTA) lightweight CNNs, FMEN and Omni-SR, the best-performing LUT method (MuLUT-X2), and HKLUT-S for generating a $3\times1280\times720$ image through $4\times$ SR. In Table~\ref{tab:memory_energy}, we demonstrate that HKLUT-S has the least storage, energy, and memory. Despite a 1dB lower PSNR in LUT compared to CNN (intrinsic to LUT schemes due to inherently restricted RF), LUT-based methods consume significantly lower energy (two orders of magnitude lower, proportional to FLOPs) and memory. This makes them ideal for low-end edge devices without specific hardware like GPU or NPU. We emphasize that our work aims to advance the Pareto front of LUT-based methods in terms of the storage-performance tradeoff. In this context, HKLUT achieves unprecedented sub-1MB results (Fig.~\ref{fig:psnr_vs_latency}).

\begin{table}[th]
    \centering
    \resizebox{1\columnwidth}{!}{\begin{tabular}{|c|cccc|}
    \hline
     Metrics   & FMEN & Omni-SR & MuLUT-X2 & HKLUT-S \\
    \hline\hline
        Storage (KB) & 769 & 792 & 4160 & \textbf{100} \\
        PSNR (dB) & 27.63 & \textbf{27.71} & 26.89 & 26.73\\
        Energy (\textmu J) & 22630 & 18432 & 701 & \textbf{117}\\
        Memory (MB) & 1024 & 6712 & 245 & \textbf{93}\\
    \hline
    \end{tabular}}
    \caption{Comparison between LUT-based approaches and lightweight CNNs. The results are obtained by performing $4\times$ SR rendering on a $1280\times720$ RGB image.}
    \label{tab:memory_energy}
\end{table}

\vspace{-4mm}

\section{Additional Ablation Studies}
\label{appendix:ablation}

\subsection{Ablation on Residual Connection} 
Table~\ref{tab:ablation_residual} compares the results of the three commonly used upsampling methods (nearest, bilinear, and bicubic). The effects of the three methods are almost indistinguishable. To reduce runtime, we selected the nearest interpolation as our upsampling strategy for residual connections.

\begin{table}[th]
\centering
\resizebox{1\columnwidth}{!}{%
\begin{tabular}{|l|ccc|}
\hline
\multirow{2}{*}{Method}  &  Set5 & Set14  & BSDS100 \\
 ~ & PSNR/SSIM & PSNR/SSIM  & PSNR/SSIM \\
\hline\hline
Nearest & 30.35/0.859 & 27.39/0.748 & 26.73/0.706 \\
Bilinear & 30.35/0.859 & 27.39/0.748 & \textbf{26.75/0.706} \\
Bicubic & \textbf{30.36/0.859} & \textbf{27.39/0.749} & \textbf{26.75/0.706} \\
\hline
\end{tabular}%
}
\caption{Ablation on different ways of implementing residual connection. ``Nearest'' is the same as HKLUT-S.} 
\label{tab:ablation_residual}
\end{table}

\vspace{-4mm}

\subsection{Ablation on Architecture Design} 
\label{appendix:ablation_architecture_design}
To further justify our design choice, we compared HKLUT with a baseline approach that simply combines two SOTA methods, namely SPLUT and MuLUT, by applying MuLUT to both MSB and LSB branches of SPLUT. Table~\ref{tab:SPMU} shows that such na\"ive combination not only requires more storage but also yields inferior performance compared to HKLUT-S. 

\begin{table}[h]
    \centering
    \resizebox{\linewidth}{!}{
    \begin{tabular}{|l|c|ccc|}
    \hline
    Method & Size &  Set5 & Set14  & BSDS100 \\
    \hline\hline
    SPLUT+MuLUT & 6144 KB & 30.19 & 27.29 & 26.70\\
    HKLUT-S & \textbf{100 KB} & \textbf{30.35} & \textbf{27.39} & \textbf{26.73} \\
    \hline
    \end{tabular}}
    \caption{HKLUT-S outperforms a piece-together of SPLUT and MuLUT with a remarkably smaller size for the 4$\times$ SR task.}
    \label{tab:SPMU}
\end{table}

\clearpage

\end{document}